\documentclass[reprint, footinbib, amsmath, amssymb, aps, prx, floatfix, twocolumn]{revtex4-2}

\usepackage{physics}
\usepackage{printlen} 
\usepackage{amsthm}
\usepackage{mathtools}

\usepackage{graphicx}
\graphicspath{{images/}}
\usepackage{wrapfig}

\usepackage{dcolumn}
\usepackage{bm}
\usepackage[hidelinks]{hyperref}
\usepackage{tikz}
\usepackage{dsfont}
\usetikzlibrary{quantikz}

\usepackage{diagbox}

\usepackage{lipsum} 

\usepackage[labelformat=simple]{subcaption}

\usepackage{ragged2e}
\DeclareCaptionJustification{justified}{\justifying}
\usepackage{printlen}

\usepackage{etoolbox}

\usepackage[english]{babel}

\usepackage[innercaption]{sidecap}
\sidecaptionvpos{figure}{c}

\usepackage{siunitx}
\AtBeginDocument{\RenewCommandCopy\qty\SI}
\makeatletter
\renewcommand{\p@subsubsection}{}
\makeatother

\newcommand{\be}{\begin{equation}} 
\newcommand{\ee}{\end{equation}}

\catcode`,\active

\catcode`\,12

\usepackage{tcolorbox}

\usepackage{environ}  
\usepackage{etoolbox} 
\usepackage{tikz}
\usetikzlibrary{arrows.meta,positioning,fit,backgrounds,calc}
\newbool{quickarxiv}
\boolfalse{quickarxiv} 
\usepackage{makecell}
\newcommand{\quickarxiv}[1]{\ifbool{quickarxiv}{}{#1}}

\begin{document}
\preprint{APS/123-QED}

\title{Beyond Unitary Quantum Simulation: Open-System Approaches to Quantum Chemistry toward Quantum Advantage}


\author{Michael Marthaler}
\affiliation{HQS Quantum Simulations GmbH, Rintheimer Str. 23, 76131 Karlsruhe, Germany}
\author{Elias Zapusek}
\affiliation{
    Institute for Quantum Electronics,
    ETH Z\"{u}rich, Otto-Stern-Weg 1, 8093 Z\"{u}rich, Switzerland
    }
\affiliation{Quantum Center, ETH Zürich, 8093 Zürich, Switzerland}
\author{Florentin Reiter}
\email{florentin.reiter@iaf.fraunhofer.de}
\affiliation{Fraunhofer Institute for Applied Solid State Physics IAF, Tullastr. 72, 79108 Freiburg, Germany}
\affiliation{
    Institute for Quantum Electronics,
    ETH Z\"{u}rich, Otto-Stern-Weg 1, 8093 Z\"{u}rich, Switzerland
    }
\affiliation{Quantum Center, ETH Zürich, 8093 Zürich, Switzerland}

\date{\today}


\begin{abstract}
Quantum simulation is widely regarded as one of the most promising routes to genuine quantum advantage, yet most existing approaches to quantum chemistry are formulated in terms of closed-system, unitary dynamics and ground-state preparation within the Born--Oppenheimer approximation. In this review, we discuss a broader perspective motivated by the observation that naturally occurring quantum systems are rarely isolated and often reach physically relevant states only through relaxation, decoherence, and thermalization. We first examine what is and is not known about exponential quantum advantage in chemistry, emphasizing that coherent Hamiltonian simulation provides the clearest formal case for speed-up, while many open questions remain for realistic problems. We then discuss how dissipation might ideally be integrated into quantum chemistry on a fault-tolerant quantum computer, using recent proposals for chemically motivated dynamical simulation as a guiding vision. More generally, we highlight the practical appeal of this approach to enhancing the robustness of quantum algorithms.
\end{abstract}


\maketitle

\section{Introduction}
\begin{figure}
    \centering
    \includegraphics{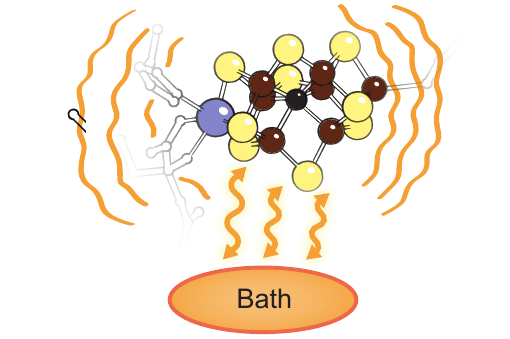}
    \caption{Molecule (FeMoco cluster) coupled to a thermal environment, illustrating the open-system setting that pervades chemistry, condensed-matter physics, and materials science. Standard quantum algorithms largely abstract away this system–bath interaction, treating the electronic problem as closed and unitary. The dissipative simulation framework discussed in this review instead takes the environmental coupling as a computational resource rather than a nuisance.}
    \label{fig:molecule}
\end{figure}

One of the central promises of quantum computing is that a controllable quantum device should be able to reproduce the behavior of physical quantum systems found in nature. In practice, however, much of the literature has implemented this idea in a restricted form: quantum algorithms are typically formulated for closed-system Hamiltonian dynamics, while naturally occurring quantum systems are almost never perfectly isolated. In chemistry, condensed-matter physics, and materials science, relevant degrees of freedom are embedded in an environment, undergo relaxation and exchange energy with surrounding modes, and are frequently observed only after some form of equilibration; a class of dynamics broadly referred to as dissipative evolution. A typical situation given by a molecule coupled to a thermal environment is illustrated in Fig. \ref{fig:molecule}. Conventional quantum chemistry also generally considers only closed systems. The central computational objects are most often the electronic ground-state energy and properties derived from it within the Born--Oppenheimer approximation. A substantial body of work has therefore also focused on quantum algorithms to calculate ground state energies. Although these algorithms almost always use a unitary time evolution or some similar unitary object as a primitive. 

At the same time, a parallel literature has developed around the simulation of heuristic open quantum models on quantum hardware. This includes spin--boson models in superconducting circuits and trapped ions, digital simulation of open-system dynamics, and more recent gate-based demonstrations of system--bath physics and dissipative many-body state preparation. In many of these works, the objective is not to simulate an isolated Hamiltonian, but to emulate the interplay between a system and its environment, often with the explicit goal of capturing relaxation, decoherence, or bath-assisted state preparation. Yet, it remains the case, that these models are often quite far removed from actual use cases in the material sciences. 

The recent work of Chen, Huang, Preskill, and Zhou \cite{ChenHuangPreskillZhou2023LocalMinima} provides a strong motivation to try to unify the work on quantum chemistry for quantum computer with the heuristics for open system dynamics. Because the work gives formal expression to an intuition that has long motivated quantum simulation heuristics: if a quantum computer is meant to reproduce the processes by which nature itself reaches physically relevant states, then dissipation and relaxation should be considered essential elements of the simulation architecture. 

At the same time, adopting this viewpoint changes the computational target in an important way. In standard quantum chemistry, the starting point is typically the electronic ground-state problem within the Born--Oppenheimer approximation, from which one derives energies, structures, and eventually dynamical or thermodynamic information. By contrast, the dissipative local-minimum framework does not directly begin with the exact ground-state energy as its primary output. Instead, it replaces the usual target by a stationary thermal state  \cite{ChenHuangPreskillZhou2023LocalMinima}. This marks a significant departure from conventional quantum chemistry. This also implies that calculations should be done directly beyond the Born--Oppenheimer approximation, since thermal equilibria should include at least vibrations. In principle we can assume that this approach
will capture all relevant properties in nature, but its implementation requires a complete reworking of the existing quantum chemistry machinery. 

In this review, we will focus only a few key points:
\begin{itemize}
    \item Can we expect exponential speed for solving problems in quantum chemistry on a quantum computer (Sec. \ref{sec:speedup})?
    \item What would be the ideal way of integrating dissipation with quantum chemistry on a quantum computer (Sec. \ref{sec:ideal})?
    \item What would be a practical way of integrating dissipation with quantum chemistry on classical and quantum computers (Sec. \ref{sec:practical})?
    \item What practical advantages does dissipation offer for quantum algorithm design (Sec. \ref{sec:algorithms})?
\end{itemize}

Finally, in Sec. \ref{sec:conclusion} we give a brief conclusion.

\section{On exponential speed up}
\label{sec:speedup}

Coherent Hamiltonian simulation remains the clearest in-principle route to quantum advantage in physics and chemistry: simulating the real-time dynamics of local quantum systems, even when the final output is only an observable such as $\langle O(t)\rangle$, is one of the foundational applications of quantum computers and is widely regarded as classically hard in the worst case \cite{Feynman1982, Lloyd1996, HamiltonianSimulationSurvey}. This observation, however, does not by itself settle the situation for practical quantum chemistry. Many chemically relevant tasks are closer to low-energy or ground-state problems than to arbitrary unitary dynamics, and these are harder even at the level of complexity theory: in general, the local Hamiltonian problem is $\mathsf{QMA}$-hard \cite{HamiltonianSimulationSurvey}. 
Moreover, even when unitary algorithms such as phase estimation are formally efficient, their practical usefulness depends on preparing initial states with sufficiently large overlap with the target state, while realistic chemistry problems may still admit strong structure-exploiting classical approximations \cite{LeeEtAl2023EvidenceExponentialAdvantageChemistry,FratusEnenkelZankerReinerMarthalerSchmitteckert2025NMR,SchleichKristensenCamposGonzalezAnguloAvaglianoBagherimehrabAldossaryGorgullaFitzsimonsAspuruGuzik2026ChemicallyMotivated}. For the purposes of the present review, the essential point is therefore simply that coherent quantum simulation provides both the strongest known case for quantum advantage and despite that fact a substantial amount of unresolved questions remain\cite{Feynman1982,Lloyd1996,HamiltonianSimulationSurvey,LeeEtAl2023EvidenceExponentialAdvantageChemistry,FratusEnenkelZankerReinerMarthalerSchmitteckert2025NMR}. Adding dissipation to the toolbox of quantum chemistry could be a way to resolve some of these questions. 

However, at first a closely related ambiguity appears again when one moves from coherent state preparation to dissipative strategies. At an intuitive level, it is tempting to believe that decoherence and dissipation should make quantum problems easier rather than harder. If a system is allowed to relax, exchange energy with its environment, and shed high-energy excitations, then one might expect that the effective region of Hilbert space relevant to the computation becomes dramatically reduced. From the perspective of classical simulation, this suggests a natural counter-hypothesis to quantum advantage: perhaps physically realistic dissipation acts as a compression mechanism, guiding the dynamics into a low-complexity manifold that can also be tracked efficiently by suitably designed classical methods. In this context the results from Chen, Huang, Preskill, and Zhou  \cite{ChenHuangPreskillZhou2023LocalMinima}
are particularly interesting. Their work proves that an exponential separation between classical and quantum computation can exist even in a framework based on thermal perturbations and dissipative local moves. In this sense, the paper rules out the naive conclusion that dissipation must automatically destroy quantum advantage. Yet the mechanism by which this separation is established also clarifies the limits of what has been shown. The proof relies on a highly structured circuit-to-Hamiltonian, or clock, construction, closely related in spirit to the constructions used in complexity-theoretic embeddings of quantum computation into Hamiltonian problems. The resulting Hamiltonians are designed so that the relevant gap scales only inverse-polynomially with system size, which is precisely what makes the separation mathematically accessible. This is an important theoretical achievement, but it is also quite far removed from the kinds of Hamiltonians that typically arise in electronic structure theory, molecular dynamics, or realistic system--bath models.

However, from the perspective of practical simulation this unresolved issue is not a weakness but an opportunity. If physically realistic dissipative ans\"atze were to retain an exponential quantum advantage, that would provide a powerful new paradigm for quantum simulation beyond the purely unitary setting. But the opposite outcome would be equally profound. If the inclusion of realistic environmental effects were to render large classes of chemically relevant quantum systems classically efficiently simulable, then this would reveal something deep about the computational structure of nature itself: namely, that the dissipative environments surrounding natural quantum systems may actively constrain them to a tiny and especially structured corner of Hilbert space.

In this sense, both outcomes are scientifically exciting. A robust quantum advantage in the presence of realistic dissipation would strengthen the case for open-system quantum simulation as a fundamentally new computational tool. A demonstration that realistic dissipation often induces practical classical tractability would be no less remarkable, because it would help explain why many naturally occurring quantum systems appear far more regular and computationally manageable than a worst-case complexity analysis would suggest. 

The analysis of quantum chemistry from an open system perspective, is a daunting task. But we feel that it is worthwhile and no matter the final judgment on quantum advantage can lead to amazing new insights. 
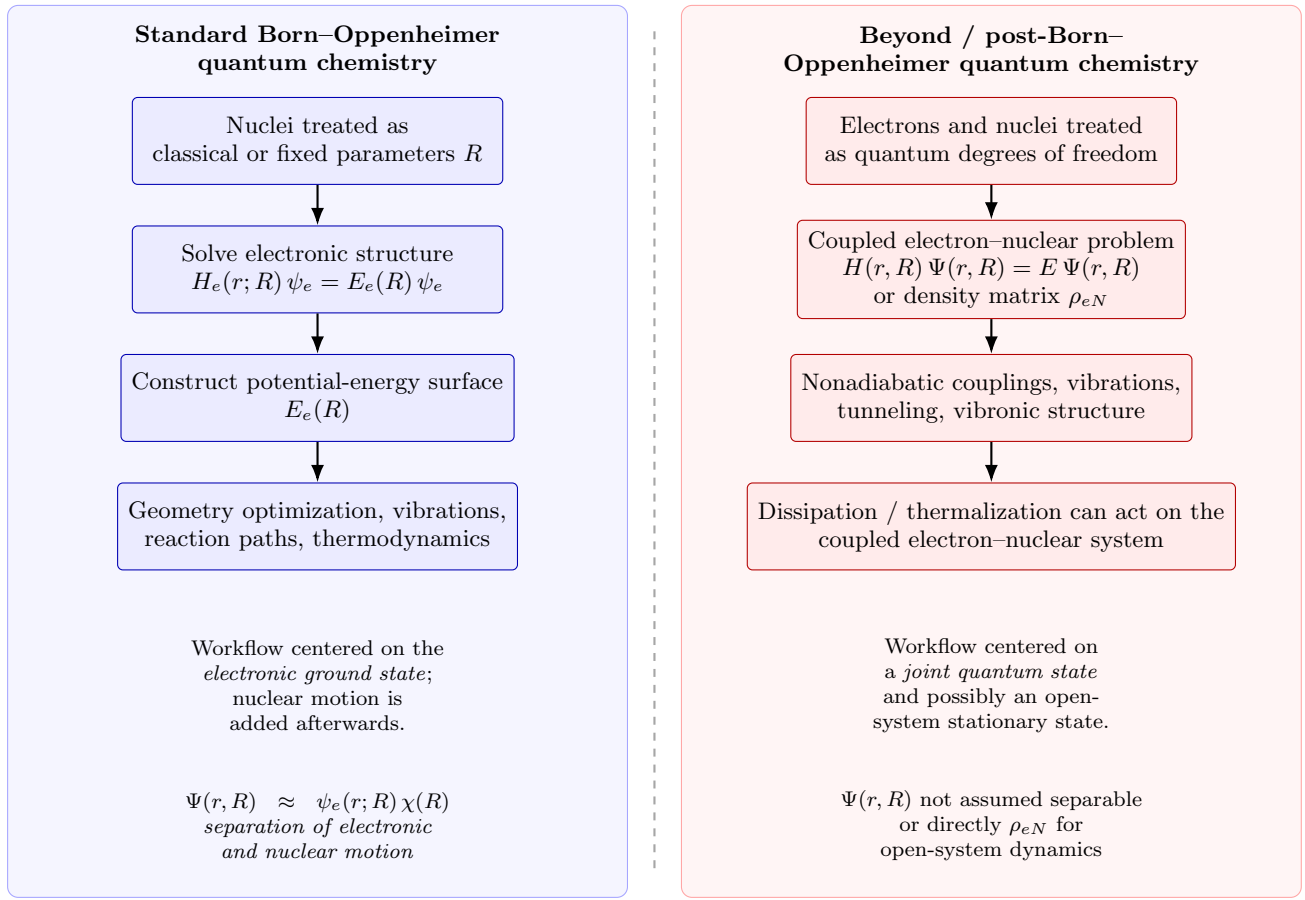
\begin{figure*}[t]
\centering
\begin{tikzpicture}[
    font=\small,
    >=Latex,
    box/.style={
        draw,
        rounded corners=2pt,
        align=center,
        minimum width=4.9cm,
        minimum height=1.15cm,
        inner sep=4pt
    },
    bluebox/.style={
        box,
        fill=blue!8,
        draw=blue!70!black
    },
    redbox/.style={
        box,
        fill=red!8,
        draw=red!70!black
    },
    titlebox/.style={
        align=center,
        font=\bfseries,
        text width=6.8cm
    },
    note/.style={
        align=center,
        font=\footnotesize,
        text width=5.6cm
    }
]

\def\panelw{8.2cm}
\def\panelh{11.8cm}
\def\gap{0.35cm}

\node[
    draw=blue!35,
    fill=blue!4,
    rounded corners=4pt,
    minimum width=\panelw,
    minimum height=\panelh,
    anchor=north east
] (Lpanel) at (-\gap,0) {};

\node[
    draw=red!35,
    fill=red!4,
    rounded corners=4pt,
    minimum width=\panelw,
    minimum height=\panelh,
    anchor=north west
] (Rpanel) at (\gap,0) {};

\draw[dashed, thick, gray!70] (0,-0.35) -- (0,-11.45);

\node[titlebox] (Ltitle) at ([yshift=-0.6cm]Lpanel.north) {Standard Born--Oppenheimer quantum chemistry};
\node[titlebox] (Rtitle) at ([yshift=-0.6cm]Rpanel.north) {Beyond / post-Born--Oppenheimer quantum chemistry};

\node[bluebox] (L1) at ([yshift=-1.8cm]Lpanel.north) {Nuclei treated as\\ classical or fixed parameters $R$};
\node[bluebox] (L2) at ([yshift=-3.5cm]Lpanel.north) {Solve electronic structure\\ $H_e(r;R)\,\psi_e = E_e(R)\,\psi_e$};
\node[bluebox] (L3) at ([yshift=-5.2cm]Lpanel.north) {Construct potential-energy surface\\ $E_e(R)$};
\node[bluebox] (L4) at ([yshift=-6.9cm]Lpanel.north) {Geometry optimization, vibrations,\\ reaction paths, thermodynamics};

\node[note] (Lnote) at ([yshift=-9.0cm]Lpanel.north)
{Workflow centered on the\\ \emph{electronic ground state};\\ nuclear motion is\\ added afterwards.};

\node[note] (Leq) at ([yshift=-10.85cm]Lpanel.north)
{$\Psi(r,R)\,\approx\,\psi_e(r;R)\,\chi(R)$\\
\emph{separation of electronic\\ and nuclear motion}};

\node[redbox] (R1) at ([yshift=-1.8cm]Rpanel.north) {Electrons and nuclei treated\\ as quantum degrees of freedom};
\node[redbox] (R2) at ([yshift=-3.5cm]Rpanel.north) {Coupled electron--nuclear problem\\ $H(r,R)\,\Psi(r,R)=E\,\Psi(r,R)$\\ or density matrix $\rho_{eN}$};
\node[redbox] (R3) at ([yshift=-5.2cm]Rpanel.north) {Nonadiabatic couplings, vibrations,\\ tunneling, vibronic structure};
\node[redbox] (R4) at ([yshift=-6.9cm]Rpanel.north) {Dissipation / thermalization can act on the\\ coupled electron--nuclear system};

\node[note] (Rnote) at ([yshift=-9.0cm]Rpanel.north)
{Workflow centered on\\ a \emph{joint quantum state}\\ and possibly an open-\\ system stationary state.};

\node[note] (Req) at ([yshift=-10.85cm]Rpanel.north)
{$\Psi(r,R)$ not assumed separable\\
or directly $\rho_{eN}$ for\\ open-system dynamics};

\draw[->, thick] (L1) -- (L2);
\draw[->, thick] (L2) -- (L3);
\draw[->, thick] (L3) -- (L4);

\draw[->, thick] (R1) -- (R2);
\draw[->, thick] (R2) -- (R3);
\draw[->, thick] (R3) -- (R4);

\end{tikzpicture}
\caption{Conceptual comparison between standard Born--Oppenheimer quantum chemistry and beyond-/post-Born--Oppenheimer approaches. In the Born--Oppenheimer picture, one first solves an electronic structure problem for fixed nuclear coordinates and then derives nuclear dynamics from the resulting potential-energy surface. In the beyond-Born--Oppenheimer picture, electrons and nuclei are treated on a common quantum footing, allowing nonadiabatic effects, vibronic structure, tunneling, and, in principle, dissipation acting on the coupled electron--nuclear system.}
\label{fig:bo_vs_postbo}
\end{figure*}

\section{What would be the ideal way of integrating dissipation with quantum chemistry on a quantum computer?}
\label{sec:ideal}

A compelling large-scale vision for this question is provided by the recent paper of Schleich \emph{at al.} on chemically motivated simulation problems \cite{SchleichKristensenCamposGonzalezAnguloAvaglianoBagherimehrabAldossaryGorgullaFitzsimonsAspuruGuzik2026ChemicallyMotivated}. The key idea in that work is to move away from formulating quantum chemistry on a quantum computer as a ground-state problem. Instead, the authors advocate a fully dynamical perspective: relatively simple atomic states are prepared first, these are then combined into molecular states through controlled scattering or assembly processes, and the resulting states are subsequently evolved in time to extract observables such as reaction probabilities, transition amplitudes, spectroscopic signals, or free-energy-related quantities. In this picture, traps, external fields, measurements, and in principle also baths and dissipative channels are  natural parts of the simulation architecture. From the perspective of the present paper, this is perhaps the clearest currently available big picture vision for how one might eventually combine quantum chemistry, dynamics, and dissipation within one unified quantum-computational framework.

At the same time, such a vision is evidently extremely ambitious. If taken literally, it asks a quantum computer to treat electrons and at least some nuclei quantum mechanically, to represent large-amplitude molecular motion, to allow for reaction pathways, and to include environment-induced relaxation or cooling in a controlled fashion. This is far beyond the scope of near-term devices and is very likely also beyond what one would want to attempt on the earliest generations of fault-tolerant hardware. It is therefore useful to ask what the scaling would look like even at the level of rough order-of-magnitude estimates.

Before discussing the most efficient first-quantized plane-wave algorithms that would suit the problem \cite{BabbushEtAl2018LowDepthMaterials,BabbushEtAl2019SublinearBasis,SuBerryWiebeRubinBabbush2021FirstQuantization}, it is useful to introduce a simpler brute-force baseline.
In Fig. \ref{fig:bo_vs_postbo}, we compare the both standard Born--Oppenheimer quantum chemistry and beyond-Born--Openheimer approaches.
For a genuinely beyond-Born--Oppenheimer problem with large-amplitude nuclear motion it is not natural to assume from the outset a compact atom-centered basis, since such bases exploit precisely the fixed-geometry structure that is absent in a general all-particle treatment. We therefore take a discretized real-space basis for our first estimate. 

We assume that we consider a problem with $A$ atoms (which for organic molecules would e.g. mean the total number of particles is or order $\eta \propto 10 A $) and that the simulation is performed in a cubic box whose linear size $L$ scales with the system size, $L \propto A$, while the spatial resolution $\Delta x$ is kept fixed. Then the number of grid points in one direction scales as $n = L/\Delta x \propto A$, and the total number of one-particle grid basis states scales as
$
N_{\mathrm{grid}} = n^3 \propto A^3.
$
A naive occupancy-based mapping of this discretized coordinate basis to a quantum computer therefore requires
\begin{equation}
    Q_{\mathrm{std}} \sim N_{\mathrm{grid}} \sim A^3
\end{equation}
qubits, up to species- and occupancy-dependent prefactors. Overall this is an extremely unfavorably scaling, requiring millions of qubits even for comparably small problems. 

However, at least in this basis, the Hamiltonian is more structured than in a generic orbital representation: the kinetic term is local on the grid, while the Coulomb interaction appears as density-density couplings between grid points. A brute-force simulation based on term-by-term application therefore leads to the rough estimate of the gate depth
\begin{equation}
G_{\mathrm{std}} \sim N_{\mathrm{grid}}^2 \sim A^6.
\end{equation}
Given that the prefactors are quite simple to estimate for this algorithmic approach, we can see that this would lead to gate depth of more than $10^{10}$ for as few as 42 atoms.  And this is the gate depth even before we add the gates necessary for implementing dissipation.

To achieve more favorable scaling we turn to the first-quantized algorithms of Babbush and co-workers \cite{BabbushEtAl2018LowDepthMaterials,BabbushEtAl2019SublinearBasis,SuBerryWiebeRubinBabbush2021FirstQuantization}. The key idea is that when the number of particles is much smaller than the number of one-particle basis functions, first quantization can compress the memory requirement dramatically. In a conventional second-quantized mapping, the number of qubits scales roughly with the number of orbitals. By contrast, in first quantization one stores the coordinates of each particle explicitly, so the qubit cost scales with the number of particles times the logarithm of the basis size. This becomes especially attractive in beyond-Born--Oppenheimer settings, where the number of quantum particles may still be moderate, but the basis required to describe both electrons and nuclei over a large reactive region can be enormous.


The Babbush first-quantized mapping changes the scaling rather dramatically. Since one stores particle coordinates rather than orbital occupation numbers, the qubit count becomes
$
Q_{\mathrm{B}} \sim \eta \log N \sim A \log A.
$
Using the first-quantized plane-wave gate estimates in their simplest form \cite{SuBerryWiebeRubinBabbush2021FirstQuantization},
$
G_{\mathrm{B}} \sim \eta^{8/3} N^{1/3},
$
one obtains
$
G_{\mathrm{B}} \sim A^{8/3} \cdot A = A^{11/3},
$
suppressing the dependence on simulation time and precision. This is the most attractive asymptotic feature of the plane-wave first-quantized approach: the number of qubits drops from cubic to essentially linear-logarithmic scaling in the number of atoms, while the gate count remains polynomial with a much smaller exponent than the naive second-quantized baseline. The important caveat is that this expression hides precisely the parameters that are likely to be most painful in an all-particle calculation. Once electrons and nuclei are treated on the same footing, the spectral range becomes broad, the relevant time scales become widely separated, and the precision requirements may become severe. Thus the favorable scaling
$
A^{11/3}
$
should be  not be read as a realistic full resource count.

The next question is how this picture changes when dissipation is added in the spirit of Chen, Huang, Preskill, and Zhou \cite{ChenHuangPreskillZhou2023LocalMinima}. Here one should distinguish between two very different notions of including dissipation. If one simply wants to simulate a Lindbladian step, the extra qubit overhead is comparatively mild: the main system register is supplemented by ancillas for block-encoding jump operators, small time- or frequency-registers, and a modest number of measurement and reset qubits. In that sense, the addition of dissipation does not radically change the memory requirements. The more demanding interpretation is to implement the full thermal-gradient-descent strategy used in the Preskill framework to guarantee convergence toward a local minimum. In that case, one pays not only for simulating dissipative steps, but also for repeatedly estimating gradients and scanning over many jump directions.

At the level of a rough scaling estimate, the qubit count remains comparatively benign. The ancilla overhead for a block-encoded Lindbladian is polylogarithmic in the number of jump operators and does not change the leading dependence on the system size. One therefore still expects, at the scaling level,
\begin{equation}
    Q_{\mathrm{B+diss}} \sim A \operatorname{polylog}(A).
\end{equation}
The gate count is less forgiving. In the thermal-gradient-descent algorithm of Chen \emph{et al.} \cite{ChenHuangPreskillZhou2023LocalMinima}, the number of dissipative update steps scales as $T=O(B^3/\varepsilon^2)$, where $B$ is an upper bound on $\|H\|_\infty$. If one makes the assumption that  $B\sim A$, and that the number of local dissipative jump operators grows linearly with system size, $m\sim A$, then the dissipative overhead scales as $mT\sim A^4$ up to precision dependence (which as discussed about can be severe). Combining this with the coherent first-quantized scaling gives the rough estimate
\begin{equation}
G_{\mathrm{B+diss}} \sim A^{23/3}.
\end{equation}
From the perspective of asymptotic computability, this is impressive. From the perspective of realistic hardware planning, however, it is implausible and certainly not a target for early fault-tolerant quantum computers.

It is useful to summarize these estimates in a compact way. In the table below, the entry \emph{standard mapping} refers to a naive second-quantized mapping of the same beyond-Born--Oppenheimer plane-wave model and is therefore best interpreted as a conservative baseline rather than an optimized algorithm.

\begin{table*}[t]
\centering
\caption{Rough asymptotic scaling estimates for an all-particle beyond-Born--Oppenheimer treatment of a chemical system with $A=N_{\mathrm{atoms}}$, under the assumptions $\eta \sim 11A$, $L \propto A$, and fixed spatial resolution $\Delta x$, so that the one-particle basis size scales as $N \sim A^3$. Time and precision dependence are omitted. }
\begin{tabular}{|l|c|c|}
\hline
Method & Scaling Gates & Scaling Qubits \\
\hline
Standard mapping to quantum computer & $A^{6}$ (brute-force baseline) & $A^3$ \\
Plain-wave first-quantized approach & $A^{11/3}$ & $A \log A$ \\
Plain-wave first-quantized approach + dissipation & $A^{23/3}$  & $A \operatorname{polylog} (A) $ \\
\hline
\end{tabular}
\label{tab:qc_dissipation_scaling}
\end{table*}

Taken at face value, Table~\ref{tab:qc_dissipation_scaling} is simultaneously encouraging and sobering. It is encouraging because it suggests that a fully quantum-mechanical treatment of electrons and nuclei together with engineered dissipation is not obviously ruled out by qubit count alone. It is sobering because the corresponding gate counts remain large, and because the omitted dependence on time and precision may be particularly severe in the beyond-Born--Oppenheimer regime. 

Finally, there is also a more conceptual limitation to the idea of simulating everything at once. It is undeniably fascinating to imagine loading as much of nature as possible into one quantum computer and letting the device reproduce the full coupled dynamics of electrons, nuclei, and environment. Yet this ideal can also be scientifically restrictive. In much of present-day quantum chemistry, a great deal of insight is gained from intermediate representations: approximate potential-energy surfaces, reaction coordinates and qualitative connections of structure to properties. These intermediate objects are  often the very language through which chemists form qualitative understanding and design follow-up calculations. A future dissipation-based quantum chemistry framework will therefore likely be most useful not when it hides all structure inside one enormous quantum evolution, but when it still allows the scientist to extract interpretable intermediate information and to combine quantum output with chemical reasoning in an iterative way.

\section{What would be a practical way of integrating dissipation with quantum chemistry on classical and quantum computers?}
\label{sec:practical}

The discussion in the previous sections has focused on the most ambitious formulation of the problem: a fully quantum-mechanical treatment of chemistry in which electronic structure, nuclear motion, and dissipative processes are all incorporated into one unified framework. While this vision is conceptually attractive, it is also clear that such an approach is extremely demanding and, for the foreseeable future, unlikely to be the most practical route toward scientific progress. A more realistic strategy is therefore to identify restricted settings in which dissipation is physically meaningful, computationally useful, and compatible with methods that can already be explored either classically or on early generations of fault-tolerant quantum computers. In this section, we outline several such paths. The unifying theme is that one should not attempt to include \emph{all} of chemistry and \emph{all} of dissipation at once, but rather choose situations in which one important aspect of open-system physics can be isolated and studied in a controlled manner.

\subsection{Targeting solid-state materials}

Perhaps the most natural practical setting is provided by solid-state materials. From the perspective of quantum algorithms, this is attractive because periodic systems are precisely the regime in which first-quantized plane-wave approaches proposed by Babbush {et al.} type are most naturally formulated. In addition, the atomic structure of many crystalline materials is already known to high precision from x-ray diffraction and related structural probes. This means that one can often begin from a fixed atomic geometry and focus primarily on the electronic structure problem, without immediately having to confront the full beyond-Born--Oppenheimer challenge.

In such a setting, dissipation can be introduced in a comparatively controlled way. One may, for example, ask how electronic excitations relax in the presence of phonons, how an optically driven electronic state thermalizes, or how electronic correlations are modified by weak coupling to an environment. From a modeling point of view, this is already a substantial simplification compared to the molecular reaction problems discussed earlier. The nuclei need not be treated as fully quantum dynamical degrees of freedom, and the periodic structure makes the use of plane-wave encodings especially natural. As a consequence, the memory advantage of first quantization can be exploited in a regime for which the basis itself is physically well matched to the problem.

This makes solid-state materials a particularly promising testing ground for dissipation-driven quantum chemistry or quantum materials simulation on a quantum computer. One can begin with well-characterized structures, focus on electronic Hamiltonians in periodic geometry, and then add open-system effects in a controlled manner. Of course the exact physical interpretation of the open system effects would be missing and this procedure would mostly serve as a way to guarantee that the algortihm converges to a resonable equilibrium.  While still difficult, this path avoids the simultaneous need to solve structural chemistry, nuclear quantum dynamics, and environment coupling all at once.

\subsection{Very small molecules as beyond-Born--Oppenheimer test beds}

At the opposite end of the spectrum lies a second highly practical direction: choosing simple molecules -- such as the diatomic molecules $H_2$ or $O_2$ -- in order to study the connection between beyond-Born--Oppenheimer physics, dissipation, and simulability at a fundamental level. Here the goal is not chemical realism, but conceptual clarity. Systems as small as two hydrogen atoms, or similarly minimal few-body models, already allow one to ask nontrivial questions about the interplay between electronic motion, nuclear motion, and environmental relaxation.

Such problems have several advantages. First, they can often be studied classically with high accuracy. This means that the scientific questions can be explored already today without waiting for large-scale quantum hardware. Second, because the systems are so small, one can vary the description systematically: one may compare Born--Oppenheimer and beyond-Born--Oppenheimer treatments, switch dissipation on and off, reduce the geometry to one spatial dimension, or vary the bath coupling in a controlled way. In this sense, very small molecules provide an ideal laboratory for developing intuition about which aspects of dissipation increase or decrease classical and quantum simulability.

A particularly appealing strategy is to start with one-dimensional or otherwise symmetry-reduced models. For example, two hydrogen atoms moving along a line already define a system in which electronic structure, bond formation, and nuclear motion can be studied together while keeping the Hilbert space sufficiently small for careful benchmarking. Since this is an obvious test case, the combination of beyond Born--Oppenheimer and electronic structure is already well studied, see e.g. Ref. \cite{YangWhite2019Diatomic}).
Here the authors study 1D diatomic molecules where both electrons and nuclei are treated as quantum particles, explicitly “going beyond the usual Born–Oppenheimer approximation”). In such setting, adding open system physics should be feasible. While such minimal models will not answer all questions of chemical relevance, they may nonetheless be invaluable for learning which dissipative mechanisms are scientifically meaningful and which algorithmic representations are actually viable.

\subsection{Quantum electrons with classical nuclear motion}

A third practical route is to combine a quantum treatment of the electrons with a classical treatment of nuclear motion. This is, in many ways, the most natural intermediate step between standard electronic-structure theory and a fully open-system beyond-Born--Oppenheimer framework. Mixed quantum-classical dynamics already plays a central role in chemical physics, especially in the description of photoexcitation, non-equilibrium electron dynamics, and light-driven structural response.

Importantly, this direction is not merely hypothetical: there already exists substantial classical software infrastructure for such calculations. A prominent example is the open-source code \emph{Octopus}, which provides a computational framework for real-time electronic dynamics, light-matter interaction, and related quantum dynamical phenomena in both finite and extended systems \cite{OctopusFramework}. This makes mixed quantum-classical dynamics an excellent starting point for the systematic inclusion of dissipation. 

 In a more speculative direction, one could also ask whether parts of such a scheme could be reformulated through quantum linear-system techniques or related subroutines. At present, however, such ideas remain challenging. Within mixed quantum-classical descriptions, important contributions to the dynamics are nonlinear, and the treatment of core repulsion is not naturally reduced to a simple linear algebra problem. Thus, while this route is very promising, it is probably most interesting to consider it first for classical computers. 

\subsection{Beyond Born--Oppenheimer physics restricted to vibrations}

A fourth and especially attractive compromise is to go beyond the Born--Oppenheimer approximation, but to restrict the non-Born--Oppenheimer physics to vibrational degrees of freedom. This is a physically meaningful simplification because many experimentally relevant manifestations of nuclear quantum effects enter through vibrations, vibronic couplings, and their interaction with light or a thermal environment. At the same time, such a restriction can keep the computational cost far below that of a fully general all-particle treatment.

This direction is particularly relevant for spectroscopy. Vibrational and vibronic structure is often directly visible in infrared, Raman, and ultrafast optical experiments, and dissipation is essential for understanding linewidths, relaxation pathways, and thermal redistribution among vibrational modes. A model that treats selected vibrational coordinates quantum mechanically, couples them to electronic structure, and allows for dissipative relaxation could therefore capture a large amount of physically relevant chemistry while avoiding the enormous overhead of a full beyond-Born--Oppenheimer simulation of all nuclei.

From a practical point of view, this approach has several advantages. The number of active non-Born--Oppenheimer coordinates can be chosen selectively. One can focus on those modes that are known experimentally or chemically to be important, rather than quantizing every nuclear degree of freedom. This keeps the number of gates under better control and also makes the interpretation of the results more transparent. In this sense, a vibrationally restricted beyond-Born--Oppenheimer treatment may offer one of the most promising near-term conceptual targets for integrating dissipation with chemically relevant quantum simulation, be it on classical or quantum computers.

\subsection{Small steps towards the big picture}

Taken together, these four directions suggest a common lesson. A practical route toward dissipation-driven quantum chemistry is unlikely to emerge from a single leap toward the fully general problem. Rather, it will probably develop through a hierarchy of increasingly realistic settings: periodic solids with fixed structure, very small molecules studied as conceptual benchmarks, mixed quantum-classical approaches in which only the electrons are treated quantum mechanically, and finally restricted beyond-Born--Oppenheimer models focused on vibrations and spectroscopy. Each of these settings allows one to ask meaningful scientific questions while keeping at least part of the problem under control. In our view, this incremental strategy is not a retreat from the larger vision, but the most credible way to make that vision scientifically productive.

\section{Algorithmic advantages of dissipative protocols}
\label{sec:algorithms}

Finally, we change our perspective from use-case-driven to bottom-up and turn to the question which practical advantages can arise from the use of dissipation in the design of quantum algorithms.

From an implementation point of view the approaches outlined above are appealing as dissipative operations can produce algorithms that are better behaved on the hardware available today. Dissipative protocols can not only make physical simulation more realistic but also make algorithms more robust and variational algorithms more trainable.
 
 Dissipative protocols for state preparation offer a qualitatively different route compared to purely unitary ones \cite{plenio_cavity-loss-induced_1999,verstraete_quantum_2009,cole_dissipative_2021,zapusek_variational_2026}. The cooling scheme of \citet{raghunandan_initialization_2020} demonstrates this concretely: by coupling the system of interest to an auxiliary bosonic bath that is periodically reset, entropy is steadily extracted from the system, thus iteratively moving it towards the target state. The same underlying logic motivates quantum digital cooling \cite{polla_quantum_2021}, in which the bosonic bath is replaced by ancilla qubits that are entangled with the system and subsequently reset, implementing a similar but discretized cooling channel. In the presence of noise, such approaches offer conceptual advantages such as preparing the desired state as a steady state as demonstrated in \cite{mi_stable_2024}. 

 The qualitative differences of dissipative algorithms have an impact on other subfields. A particularly active and promising direction has emerged in which dissipative operations are leveraged as a tool to overcome one of the central trainability challenges in variational quantum algorithms: the barren plateau phenomenon \cite{singkanipa_beyond_2025,mele_noise-induced_2024, sannia_engineered_2024,zapusek_scaling_2025}. A cost function $C$ exhibits a barren plateau if the variance of its gradients is exponentially suppressed with the number of qubits $n$. Barren plateaus can arise from several sources, including excessive circuit expressiveness \cite{mcclean_barren_2018,holmes_connecting_2022}, global cost functions \cite{cerezo_cost_2021}, and high entanglement in the circuit \cite{ortiz_marrero_entanglement-induced_2021}. 
    Dissipation is emerging as a central tool to avoid them.
    
Following this direction, the authors \citet{sannia_engineered_2024} show that dissipation can be used to transform a global cost function into a local cost function. In conjunction with a quantum circuit of depth scaling at most logarithmically with system size this allows for scalable training. 

Unital noise, which brings the system state closer to the fully mixed state, significantly worsens the barren plateau problem. In the presence of unital contractive noise both gradients and expectation values are deterministically concentrated in the depth of the circuit, referred to as noise-induced barren plateaus \cite{wang_noise-induced_2021,schumann_emergence_2024,singkanipa_beyond_2025}. One can view this as a consequence of unital noise only decreasing the purity of a quantum state, exponentially approaching that of the fully mixed state which is of no use when calculating gradients \cite{singkanipa_beyond_2025}. In this setting, only dissipative, specifically non-unital operations can increase the purity of the state. These nonunitary operations simultaneously keep the average entanglement of the circuit's output state in check avoiding unitary barren plateaus. Therefore, employing nonunitary operations such as qubit reset allows for nonzero gradients even for deep circuits \cite{mele_noise-induced_2024,zapusek_scaling_2025}. 

Taken together, these results frame dissipation as an active resource rather than being an obstacle to quantum computation. It emerges that open-system behavior can be engineered to make quantum algorithms more robust, more trainable, and more naturally suited to the noisy hardware available today.

\section{Conclusion}
\label{sec:conclusion}

Quantum chemistry is considered one of the most important use-cases of quantum computing both in the near-term and in the long run.
In this review, we have addressed whether dissipation makes it harder or easier to classically simulate quantum chemistry problems. In our discussion, we have considered a unified framework where electronic structure, nuclear motion, and dissipative processes are all treated quantum-mechanically. To be concrete, we have outlined several candidate systems in which important aspects of quantum chemistry can be isolated and studied in a controlled manner, in particular fully quantum-mechanical models of electrons and nuclei, as well as the simulation of solids which is relevant to x-ray diffraction.
Taking a bottom-up approach, we broaden the scope to the construction of quantum algorithms with added dissipation. These may offer crucial advantages such as scalable training. 
For quantum chemistry, we find that the scaling of the number of gates required to reach quantum advantage for the simulation of a realistic problem remains challenging. Nonetheless, this provides guidance for future improvements in quantum hardware and algorithms along the path toward quantum advantage. Whether and when such advantages will be realized remains to be seen, as our current understanding is likely to evolve in the coming years.

\bibliography{references}

@article{zapusek_variational_2026,
	title = {Variational quantum thermalizers based on weakly-symmetric nonunitary multi-qubit operations},
	volume = {11},
	issn = {2058-9565},
	url = {https://iopscience.iop.org/article/10.1088/2058-9565/ae2886},
	doi = {10.1088/2058-9565/ae2886},
	number = {2},
	urldate = {2026-03-23},
	journal = {Quantum Science and Technology},
	author = {Zapusek, Elias and Kirova, Kristina and Hahn, Walter and Marthaler, Michael and Reiter, Florentin},
	month = jun,
	year = {2026},
	pages = {025006},
}

@article{plenio_cavity-loss-induced_1999,
	title = {Cavity-loss-induced generation of entangled atoms},
	volume = {59},
	copyright = {http://link.aps.org/licenses/aps-default-license},
	issn = {1050-2947, 1094-1622},
	url = {https://link.aps.org/doi/10.1103/PhysRevA.59.2468},
	doi = {10.1103/PhysRevA.59.2468},
	language = {english},
	number = {3},
	urldate = {2025-07-17},
	journal = {Physical Review A},
	author = {Plenio, M. B. and Huelga, S. F. and Beige, A. and Knight, P. L.},
	month = mar,
	year = {1999},
	pages = {2468--2475},
}

@article{ortiz_marrero_entanglement-induced_2021,
	title = {Entanglement-{Induced} {Barren} {Plateaus}},
	volume = {2},
	issn = {2691-3399},
	url = {https://link.aps.org/doi/10.1103/PRXQuantum.2.040316},
	doi = {10.1103/PRXQuantum.2.040316},
	language = {english},
	number = {4},
	urldate = {2025-05-22},
	journal = {PRX Quantum},
	author = {Ortiz Marrero, Carlos and Kieferova, Maria and Wiebe, Nathan},
	month = oct,
	year = {2021},
	pages = {040316},
}

@article{singkanipa_beyond_2025,
	title = {Beyond unital noise in variational quantum algorithms: noise-induced barren plateaus and limit sets},
	volume = {9},
	issn = {2521-327X},
	shorttitle = {Beyond unital noise in variational quantum algorithms},
	url = {https://quantum-journal.org/papers/q-2025-01-30-1617/},
	doi = {10.22331/q-2025-01-30-1617},
	language = {english},
	urldate = {2025-05-28},
	journal = {Quantum},
	author = {Singkanipa, Phattharaporn and Lidar, Daniel A.},
	month = jan,
	year = {2025},
	pages = {1617},
}

@article{sannia_engineered_2024,
	title = {Engineered dissipation to mitigate barren plateaus},
	volume = {10},
	issn = {2056-6387},
	url = {https://www.nature.com/articles/s41534-024-00875-0},
	doi = {10.1038/s41534-024-00875-0},
	language = {english},
	number = {1},
	urldate = {2025-05-13},
	journal = {npj Quantum Information},
	author = {Sannia, Antonio and Tacchino, Francesco and Tavernelli, Ivano and Giorgi, Gian Luca and Zambrini, Roberta},
	month = sep,
	year = {2024},
	pages = {81},
}

@article{holmes_connecting_2022,
	title = {Connecting {Ansatz} {Expressibility} to {Gradient} {Magnitudes} and {Barren} {Plateaus}},
	volume = {3},
	issn = {2691-3399},
	url = {https://link.aps.org/doi/10.1103/PRXQuantum.3.010313},
	doi = {10.1103/PRXQuantum.3.010313},
	language = {english},
	number = {1},
	urldate = {2022-09-15},
	journal = {PRX Quantum},
	author = {Holmes, Zoe and Sharma, Kunal and Cerezo, M. and Coles, Patrick J.},
	month = jan,
	year = {2022},
	pages = {010313},
}

@article{cerezo_cost_2021,
	title = {Cost function dependent barren plateaus in shallow parametrized quantum circuits},
	volume = {12},
	issn = {2041-1723},
	url = {https://www.nature.com/articles/s41467-021-21728-w},
	doi = {10.1038/s41467-021-21728-w},
	language = {english},
	number = {1},
	urldate = {2025-05-13},
	journal = {Nature Communications},
	author = {Cerezo, M. and Sone, Akira and Volkoff, Tyler and Cincio, Lukasz and Coles, Patrick J.},
	month = mar,
	year = {2021},
	pages = {1791},
}

@article{schumann_emergence_2024,
	title = {Emergence of noise-induced barren plateaus in arbitrary layered noise models},
	volume = {9},
	issn = {2058-9565},
	url = {https://iopscience.iop.org/article/10.1088/2058-9565/ad6285},
	doi = {10.1088/2058-9565/ad6285},
	number = {4},
	urldate = {2025-05-19},
	journal = {Quantum Science and Technology},
	author = {Schumann, M and Wilhelm, F K and Ciani, A},
	month = oct,
	year = {2024},
	pages = {045019},
}

@article{mi_stable_2024,
	title = {Stable quantum-correlated many-body states through engineered dissipation},
	volume = {383},
	issn = {0036-8075, 1095-9203},
	url = {https://www.science.org/doi/10.1126/science.adh9932},
	doi = {10.1126/science.adh9932},
	language = {english},
	number = {6689},
	urldate = {2024-06-05},
	journal = {Science},
	author = {Mi, X. and Michailidis, A. A. and Shabani, S. and Miao, K. C. and Klimov, P. V. and Lloyd, J. and Rosenberg, E. and Acharya, R. and Aleiner, I. and Andersen, T. I. and Ansmann, M. and Arute, F. and others and Bovaird, J. and Brill, L. and Broughton, M. and Buckley, B. B. and Buell, D. A. and Burger, T. and Burkett, B. and Bushnell, N. and Chen, Z. and Chiaro, B. and Chik, D. and Chou, C. and Cogan, J. and Collins, R. and Conner, P. and Courtney, W. and Crook, A. L. and Curtin, B. and Dau, A. G. and Debroy, D. M. and Del Toro Barba, A. and Demura, S. and Di Paolo, A. and Drozdov, I. K. and Dunsworth, A. and Erickson, C. and Faoro, L. and Farhi, E. and Fatemi, R. and Ferreira, V. S. and Burgos, L. F. and Forati, E. and Fowler, A. G. and Foxen, B. and Genois, E. and Giang, W. and Gidney, C. and Gilboa, D. and Giustina, M. and Gosula, R. and Gross, J. A. and Habegger, S. and Hamilton, M. C. and Hansen, M. and Harrigan, M. P. and Harrington, S. D. and Heu, P. and Hoffmann, M. R. and Hong, S. and Huang, T. and Huff, A. and Huggins, W. J. and Ioffe, L. B. and Isakov, S. V. and Iveland, J. and Jeffrey, E. and Jiang, Z. and Jones, C. and Juhas, P. and Kafri, D. and Kechedzhi, K. and Khattar, T. and Khezri, M. and Kieferova, M. and Kitaev, A. and Klots, A. R. and Korotkov, A. N. and Kostritsa, F. and Kreikebaum, J. M. and Landhuis, D. and Laptev, P. and Lau, K.-M. and Laws, L. and Lee, J. and Lee, K. W. and Lensky, Y. D. and Lester, B. J. and Lill, A. T. and Liu, W. and Locharla, A. and Malone, F. D. and Martin, O. and McClean, J. R. and McEwen, M. and Mieszala, A. and Montazeri, S. and Morvan, A. and Movassagh, R. and Mruczkiewicz, W. and Neeley, M. and Neill, C. and Nersisyan, A. and Newman, M. and Ng, J. H. and Nguyen, A. and Nguyen, M. and Niu, M. Y. and O'Brien, T. E. and Opremcak, A. and Petukhov, A. and Potter, R. and Pryadko, L. P. and Quintana, C. and Rocque, C. and Rubin, N. C. and Saei, N. and Sank, D. and Sankaragomathi, K. and Satzinger, K. J. and Schurkus, H. F. and Schuster, C. and Shearn, M. J. and Shorter, A. and Shutty, N. and Shvarts, V. and Skruzny, J. and Smith, W. C. and Somma, R. and Sterling, G. and Strain, D. and Szalay, M. and Torres, A. and Vidal, G. and Villalonga, B. and Heidweiller, C. V. and White, T. and Woo, B. W. K. and Xing, C. and Yao, Z. J. and Yeh, P. and Yoo, J. and Young, G. and Zalcman, A. and Zhang, Y. and Zhu, N. and Zobrist, N. and Neven, H. and Babbush, R. and Bacon, D. and Boixo, S. and Hilton, J. and Lucero, E. and Megrant, A. and Kelly, J. and Chen, Y. and Roushan, P. and Smelyanskiy, V. and Abanin, D. A.},
	month = mar,
	year = {2024},
	pages = {1332--1337},
}

@article{wang_noise-induced_2021,
	title = {Noise-induced barren plateaus in variational quantum algorithms},
	volume = {12},
	issn = {2041-1723},
	url = {https://www.nature.com/articles/s41467-021-27045-6},
	doi = {10.1038/s41467-021-27045-6},
	language = {english},
	number = {1},
	urldate = {2022-10-20},
	journal = {Nature Communications},
	author = {Wang, Samson and Fontana, Enrico and Cerezo, M. and Sharma, Kunal and Sone, Akira and Cincio, Lukasz and Coles, Patrick J.},
	month = dec,
	year = {2021},
	pages = {6961},
}

@article{polla_quantum_2021,
	title = {Quantum digital cooling},
	volume = {104},
	issn = {2469-9926, 2469-9934},
	url = {https://link.aps.org/doi/10.1103/PhysRevA.104.012414},
	doi = {10.1103/PhysRevA.104.012414},
	language = {english},
	number = {1},
	urldate = {2024-06-17},
	journal = {Physical Review A},
	author = {Polla, Stefano and Herasymenko, Yaroslav and O'Brien, Thomas E.},
	month = jul,
	year = {2021},
	pages = {012414},
}

@article{verstraete_quantum_2009,
	title = {Quantum computation and quantum-state engineering driven by dissipation},
	volume = {5},
	issn = {1745-2473, 1745-2481},
	url = {http://www.nature.com/articles/nphys1342},
	doi = {10.1038/nphys1342},
	language = {english},
	number = {9},
	urldate = {2022-07-01},
	journal = {Nature Physics},
	author = {Verstraete, Frank and Wolf, Michael M. and Ignacio Cirac, J.},
	month = sep,
	year = {2009},
	pages = {633--636},
}

@article{raghunandan_initialization_2020,
	title = {Initialization of quantum simulators by sympathetic cooling},
	volume = {6},
	issn = {2375-2548},
	url = {https://www.science.org/doi/10.1126/sciadv.aaw9268},
	doi = {10.1126/sciadv.aaw9268},
	language = {english},
	number = {10},
	urldate = {2023-05-26},
	journal = {Science Advances},
	author = {Raghunandan, Meghana and Wolf, Fabian and Ospelkaus, Christian and Schmidt, Piet O. and Weimer, Hendrik},
	month = mar,
	year = {2020},
	pages = {eaaw9268},
}

@article{mcclean_barren_2018,
	title = {Barren plateaus in quantum neural network training landscapes},
	volume = {9},
	issn = {2041-1723},
	doi = {10.1038/s41467-018-07090-4},
	number = {1},
	urldate = {2021-10-14},
	journal = {Nature Communications},
	author = {McClean, Jarrod R. and Boixo, Sergio and Smelyanskiy, Vadim N. and Babbush, Ryan and Neven, Hartmut},
	month = dec,
	year = {2018},
	pages = {4812},
}

@article{cole_dissipative_2021,
	title = {Dissipative preparation of {W} states in trapped ion systems},
	volume = {23},
	issn = {1367-2630},
	url = {https://iopscience.iop.org/article/10.1088/1367-2630/ac09c8},
	doi = {10.1088/1367-2630/ac09c8},
	number = {7},
	urldate = {2021-09-22},
	journal = {New Journal of Physics},
	author = {Cole, Daniel C and Wu, Jenny J and Erickson, Stephen D and Hou, Pan-Yu and Wilson, Andrew C and Leibfried, Dietrich and Reiter, Florentin},
	month = jul,
	year = {2021},
	pages = {073001},
}

@article{ChenHuangPreskillZhou2023LocalMinima,
  author  = {Chen, Chi-Fang and Huang, Hsin-Yuan and Preskill, John and Zhou, Leo},
  title   = {Local minima in quantum systems},
  journal = {Nature Physics},
  volume  = {21},
  pages   = {654--660},
  year    = {2025},
  doi     = {10.1038/s41567-025-02781-4}
}

@article{Feynman1982,
  author  = {Feynman, Richard P.},
  title   = {Simulating physics with computers},
  journal = {International Journal of Theoretical Physics},
  volume  = {21},
  number  = {6--7},
  pages   = {467--488},
  year    = {1982},
  doi     = {10.1007/BF02650179}
}

@article{Lloyd1996,
  author  = {Lloyd, Seth},
  title   = {Universal Quantum Simulators},
  journal = {Science},
  volume  = {273},
  number  = {5278},
  pages   = {1073--1078},
  year    = {1996},
  doi     = {10.1126/science.273.5278.1073}
}

@book{HamiltonianSimulationSurvey,
  author    = {Dalzell, Alexander M. and McArdle, Sam and Berta, Mario and Bienias, Przemyslaw and Chen, Chi-Fang and Gily{\'e}n, Andr{\'a}s and Hann, Connor T. and Kastoryano, Michael J. and Khabiboulline, Emil T. and Kubica, Aleksander and Salton, Grant and Wang, Samson and Brand{\~a}o, Fernando G. S. L.},
  title     = {Quantum Algorithms: A Survey of Applications and End-to-End Complexities},
  publisher = {Cambridge University Press},
  year      = {2025},
  doi       = {10.1017/9781009639651}
}

@misc{FratusEnenkelZankerReinerMarthalerSchmitteckert2025NMR,
  author        = {Fratus, Keith R. and Enenkel, Nicklas and Zanker, Sebastian and Reiner, Jan-Michael and Marthaler, Michael and Schmitteckert, Peter},
  title         = {Can a Quantum Computer Simulate Nuclear Magnetic Resonance Spectra Better than a Classical One?},
  year          = {2025},
  eprint        = {2508.06448},
  archivePrefix = {arXiv},
  doi           = {10.48550/arXiv.2508.06448}
}

@article{LeeEtAl2023EvidenceExponentialAdvantageChemistry,
  author  = {Lee, Seunghoon and Lee, Joonho and Zhai, Huanchen and Tong, Yu and Dalzell, Alexander M. and Kumar, Ashutosh and Helms, Phillip and Gray, Johnnie and Cui, Zhi-Hao and Liu, Wenyuan and Kastoryano, Michael and Babbush, Ryan and Preskill, John and Reichman, David R. and Campbell, Earl T. and Valeev, Edward F. and Lin, Lin and Chan, Garnet Kin-Lic},
  title   = {Evaluating the evidence for exponential quantum advantage in ground-state quantum chemistry},
  journal = {Nature Communications},
  volume  = {14},
  pages   = {1952},
  year    = {2023},
  doi     = {10.1038/s41467-023-37587-6}
}

@article{SchleichKristensenCamposGonzalezAnguloAvaglianoBagherimehrabAldossaryGorgullaFitzsimonsAspuruGuzik2026ChemicallyMotivated,
  author  = {Schleich, Philipp and Kristensen, Lasse Bj{\o}rn and Campos Gonzalez Angulo, Jorge A. and Avagliano, Davide and Bagherimehrab, Mohsen and Aldossary, Abdulrahman and Gorgulla, Christoph and Fitzsimons, Joe and Aspuru-Guzik, Al{\'a}n},
  title   = {Chemically Motivated Simulation Problems are Efficiently Solvable by a Quantum Computer},
  journal = {Digital Discovery},
  volume  = {5},
  pages   = {64--87},
  year    = {2026},
  doi     = {10.1039/D5DD00377F}
}

@article{BabbushEtAl2018LowDepthMaterials,
  author  = {Babbush, Ryan and Wiebe, Nathan and McClean, Jarrod and McClain, James and Neven, Hartmut and Chan, Garnet Kin-Lic},
  title   = {Low-Depth Quantum Simulation of Materials},
  journal = {Physical Review X},
  volume  = {8},
  pages   = {011044},
  year    = {2018},
  doi     = {10.1103/PhysRevX.8.011044}
}

@article{BabbushEtAl2019SublinearBasis,
  author  = {Babbush, Ryan and Berry, Dominic W. and McClean, Jarrod R. and Neven, Hartmut},
  title   = {Quantum Simulation of Chemistry with Sublinear Scaling in Basis Size},
  journal = {npj Quantum Information},
  volume  = {5},
  pages   = {92},
  year    = {2019},
  doi     = {10.1038/s41534-019-0199-y}
}

@article{SuBerryWiebeRubinBabbush2021FirstQuantization,
  author  = {Su, Yuan and Berry, Dominic W. and Wiebe, Nathan and Rubin, Nicholas and Babbush, Ryan},
  title   = {Fault-Tolerant Quantum Simulations of Chemistry in First Quantization},
  journal = {PRX Quantum},
  volume  = {2},
  pages   = {040332},
  year    = {2021},
  doi     = {10.1103/PRXQuantum.2.040332}
}

@article{OctopusFramework,
  author  = {Tancogne-Dejean, Nicolas and others},
  title   = {Octopus, a computational framework for exploring light-driven phenomena and quantum dynamics in extended and finite systems},
  journal = {The Journal of Chemical Physics},
  volume  = {152},
  pages   = {124119},
  year    = {2020},
  doi     = {10.1063/1.5142502}
}

@misc{mele_noise-induced_2024,
  author        = {Mele, Antonio Anna and Angrisani, Armando and Ghosh, Soumik and Khatri, Sumeet and Eisert, Jens and Stilck Fran{\c{c}}a, Daniel and Quek, Yihui},
  title         = {Noise-induced shallow circuits and absence of barren plateaus},
  year          = {2024},
  eprint        = {2403.13927},
  archivePrefix = {arXiv},
  doi           = {10.48550/arXiv.2403.13927}
}

@misc{zapusek_scaling_2025,
  author        = {Zapusek, Elias and Rojkov, Ivan and Reiter, Florentin},
  title         = {Scaling Quantum Algorithms via Dissipation: Avoiding Barren Plateaus},
  year          = {2025},
  eprint        = {2507.02043},
  archivePrefix = {arXiv},
  doi           = {10.48550/arXiv.2507.02043}
}

@article{YangWhite2019Diatomic,
  author  = {Yang, Mingru and White, Steven R.},
  title   = {Density-matrix-renormalization-group study of a one-dimensional diatomic molecule beyond the Born-Oppenheimer approximation},
  journal = {Physical Review A},
  volume  = {99},
  pages   = {022509},
  year    = {2019},
  doi     = {10.1103/PhysRevA.99.022509}
}

\end{document}